\documentclass[11pt]{article}

\usepackage[margin=1in]{geometry}
\usepackage{amsmath,amssymb,amsfonts}
\usepackage{amsthm}
\usepackage{algorithm,algorithmic}
\usepackage{bbm}
\usepackage{multirow}
\usepackage{graphicx}
\usepackage{subcaption}
\usepackage{booktabs}
\usepackage{hyperref}

\usepackage{algorithm,algorithmic}

\usepackage{bbm}

\usepackage{multirow}
\usepackage{graphicx}          

\usepackage{subcaption}

\newtheorem{lemma}{Lemma}
\newtheorem{assumption}{Assumption}

\title{Online Ensemble Learning for Sector Rotation: A Gradient-Free Framework}

\author{
  Jiaju Miao\thanks{Department of Applied Mathematics and Statistics, Stony Brook University, Stony Brook, NY, USA. Email: \texttt{miaojiaju@gmail.com}}
  \and
  Pawe\l\ Polak\thanks{Department of Applied Mathematics and Statistics, Stony Brook University, Stony Brook, NY, USA. Email: \texttt{pawel.polak@stonybrook.edu}}
}

\date{}

\begin{document}

\maketitle

\begin{abstract}
We propose a gradient-free online ensemble learning algorithm that dynamically combines forecasts from a heterogeneous set of machine learning models based on their recent predictive performance, measured by out-of-sample $R^2$. The ensemble is model-agnostic, requires no gradient access, and is designed for sequential forecasting under nonstationarity. It adaptively reweights 16 constituent models: three linear benchmarks—Ordinary Least Squares (OLS), Principal Component Regression (PCR), and LASSO—and thirteen nonlinear machine learning models, including Random Forests, Gradient-Boosted Regression Trees, and a hierarchy of feedforward neural networks (NN1–NN12).

We apply this framework to the sector rotation problem, using sector-level features derived by aggregating firm-specific characteristics. Empirically, we find that sector-level returns are more predictable and stable than individual asset returns, making them well-suited for cross-sectional forecasting. To exploit this structure, our algorithm constructs sector-specific ensembles that assign adaptive weights to constituent models in a rolling-window fashion, guided by their forecast accuracy.

Our key theoretical contribution is to bound the online forecast regret directly in terms of realized out-of-sample $R^2$, a standard empirical performance metric that here serves as the loss function in the ensemble procedure. This provides a novel and interpretable guarantee: the ensemble performs nearly as well as the best model in hindsight in terms of predictive power.

Empirical results show that the ensemble consistently outperforms individual models, equal-weighted combinations, and traditional offline ensemble methods in both predictive accuracy and economic value. When used to construct sector rotation portfolios, it delivers substantial improvements in risk-adjusted returns, maintains robustness across macroeconomic regimes, and demonstrates resilience during periods of financial stress, including the COVID-19 crisis.
\end{abstract}


\section{Introduction}
Machine learning (ML) has become integral to quantitative finance for its ability to capture nonlinear dependencies, scale to high-dimensional data, and automate forecasting. In asset pricing, ML methods extract predictive signals from firm characteristics \cite{guxiu:20, Philip:20, Gianluca:20}, often outperforming classical models such as the CAPM. Yet, individual ML models can be unstable in nonstationary environments or when optimized purely for in-sample accuracy without adequate regularization \cite{Albuquerque:22, Vapnik:00}. This motivates ensemble approaches, which combine models to reduce variance, enhance generalization, and exploit model diversity \cite{Wood:23}.

Diversity among base models has emerged as a critical determinant of ensemble success \cite{Wood:23}. Forecasting competitions, including the M5 Challenge, show that heterogeneous ensembles consistently outperform single models \cite{Vogk:22}. Industry systems such as AdaNet \cite{adanet} and Microsoft Azure AutoML \cite{sawyers:21} similarly integrate predictions across diverse black-box models.

We propose a gradient-free online ensemble algorithm tailored for financial time-series prediction. Based on the Multiplicative Weights Update Method (MWUM) \cite{alabdulmohsin:18, arora:05, bailey:18, plotkin:95}, it updates model weights dynamically from recent forecasting performance without retraining or accessing internal model details. This makes it practical for aggregating forecasts from heterogeneous sources—independent managers, teams, or organizational units—while preserving confidentiality and reducing integration complexity, ideal for decentralized or privacy-preserving infrastructures such as microprediction networks \cite{Cotton2022}.

The method employs a \textit{gain function} balancing \textit{exploitation} (rewarding accuracy) and \textit{exploration} (promoting diversity), adapting to structural changes and nonstationarity. The gain is defined via out-of-sample performance $R^2_{\text{oos}}$, and we derive probabilistic regret bounds showing that cumulative forecast error remains close to that of the best model in hindsight. 

Empirically, we test the method in sector-level return prediction. Unlike traditional sector rotation strategies relying on macro indicators \cite{Ali:21}, we aggregate firm-level features using probabilistic PCA and feed them to diverse ML models—linear, tree-based, and deep networks—whose outputs are combined by the online ensemble. This aggregation enhances the predictive signal-to-noise ratio.

Sector rotation offers both economic and practical relevance. It reallocates capital among industries based on expected performance, capturing cyclical shifts in growth, sentiment, and policy. Sector-level returns exhibit stronger persistence and lower idiosyncratic noise than firm-level returns, reflecting systematic exposures to macroeconomic factors. These properties make sector rotation central to investment vehicles such as ETFs, thematic funds, and tactical allocation mandates.

Our empirical results show that the online ensemble achieves higher out-of-sample $R^2_{\text{oos}}$ than individual models, simple averages, or offline ensembles such as stacked regression. The resulting sector rotation strategy delivers significant alphas relative to benchmark factor models \cite{Sharpe:64, Carhart:97}, remains robust across subsamples and regimes, and performs well even after accounting for transaction costs.

Overall, this study contributes to ML-based asset pricing research \cite{guxiu:20, leippold:20, Weigand:19, AndrewVan:20, Gianluca:20, Chen:21} by introducing a theoretically grounded, practical ensemble framework. It offers interpretable performance attribution, adapts to evolving predictive environments, and scales efficiently for institutional asset management.

\section{Modeling Sector Returns}\label{sec:Methodology}

We organize assets into sectors via the first two digits of their Standard Industrial Classification (SIC) codes.\footnote{We use SIC rather than newer schemes such as GICS or ICB because it provides consistent historical coverage across the full sample, aligns with regulatory datasets (e.g.\ SEC filings), and avoids the classification drift and backfill bias introduced by periodic GICS revisions over long horizons \cite{Papenkov:23, MIS2:24}.} Let \(r_{i,t+1}\) denote the excess return of sector \(i\) from time \(t\) to \(t+1\), for \(i = 1,\dots,N\) and \(t = 1,\dots,T\). We compute two sector-level return measures: (i) the equally weighted average, and (ii) the capitalization-weighted average:
\begin{equation}\label{eq:def_sector_returns}
r_{i, t+1} = 
\begin{cases}
\displaystyle \frac{1}{n_i} \sum_{j=1}^{n_i} r_{i, t+1}^{j}, & \text{(equally weighted)} \\[1em]
\displaystyle \frac{\sum_{j=1}^{n_i} c_{i, t}^{j} \, r_{i, t+1}^{j}}{\sum_{j=1}^{n_i} c_{i, t}^{j}}, & \text{(capitalization weighted)}
\end{cases}
\end{equation}
Here \(r_{i,t+1}^j\) is the return of stock \(j\) in sector \(i\), \(c_{i,t}^j\) is its market capitalization at time \(t\), and \(n_i\) is the number of firms in sector \(i\).

We posit the following additive error model for sector returns:
\begin{equation}\label{eq:r=Er+eps}
r_{i,t+1} = E_{t}(r_{i,t+1}) + \epsilon_{i,t+1}, \qquad E_{t}(r_{i,t+1}) = g(z_{i,t}),
\end{equation}
where \(z_{i,t}\in\mathbb{R}^P\) is a vector of sector-level predictors, and \(g(\cdot)\) is a (potentially nonlinear) mapping estimated separately for each sector.

The sector‐level model offers a useful balance between macroeconomic and firm‐level modeling. It retains cross‐sectional structure in firm fundamentals while averaging out idiosyncratic noise, thereby generating more stable and interpretable predictors of aggregate behavior. Sector aggregates reflect coordinated responses of firms facing similar demand, technological, and regulatory forces, and align naturally with business cycle dynamics. Empirical work shows that industry portfolios display predictability using lagged cross-industry signals \cite{RapachStraussTuZhou:2018, RapachInterdependencies:2015}, supporting the notion that sector-level modeling bridges micro and macro sources of return variation.

We opt to aggregate firm-level characteristics rather than rely solely on macroeconomic variables for several reasons. First, macro indicators are published infrequently and subject to reporting lags and revisions, which limits their usefulness for short‐horizon forecasts or high‐frequency portfolio allocation. In contrast, firm accounting and market data are more timely and often embody forward-looking expectations, making aggregated signals more responsive to shifting fundamentals \cite{BabaYara:2020}.

Second, firm-level metrics—such as valuation ratios, profitability measures, or investment behavior—typically adjust more rapidly to changes in economic or policy regimes than macro aggregates, which by construction involve delays and smoothing. Hence, sector‐level aggregates of these signals can act as leading indicators for cyclical turning points.

Third, aggregation filters out idiosyncratic noise while retaining systematic variation common across firms in a sector—such as regulatory shifts, supply chain realignments, or sectoral technological change. This filtering stabilizes the predictive signal and mitigates overfitting risk, producing inputs that combine the responsiveness of micro data with the robustness of more aggregate measures—qualities that are well suited for machine learning models targeting sector rotation and return forecasting.

To form \( z_{i,t} \), we begin with 94 firm-level characteristics collected for roughly 9,000 firms across 60 sectors (following \cite{guxiu:20}). Directly including all raw firm-level variables would yield on the order of 14,100 predictors per sector, which is untenable given sample size and overfitting risk. Therefore, we apply probabilistic principal component analysis (PPCA) separately to each characteristic–sector rolling window. PPCA extends classical PCA by embedding it in a latent variable generative model and estimating via an expectation–maximization algorithm, allowing it to handle missing data carefully \cite{TippingBishop:99}. The first principal component from each characteristic is retained to form the vector \( z_{i,t} \). Importantly, we found in empirical testing that including more than the first PC typically degraded predictive performance, likely due to overfitting and noise amplification. We adopt this parsimonious representation—one PC per characteristic—to stabilize the predictor space and focus our methodological emphasis on the ensemble learning architecture, which constitutes the main contribution of this work.

The PPCA preprocessing procedure is summarized in the following algorithm.

\begin{algorithm}[h!]
\renewcommand{\thealgorithm}{}
\begin{algorithmic}[1]
\FOR{$t=1$ to $\tau$, and $i=1$ to $N$}
    \FOR{$k=1$ to $K$}
        \STATE Apply PPCA to the matrix 
        \[
            X_{i,t}^{(k)} = \big[x_{i,n}^{j(k)}\big]_{n \in n_t,\; j=1,\dots,n_i}
        \]
        of dimension $|n_t| \times n_i$.\\
        \STATE Extract the first principal component (PC1): $f_{i,t}^{(k)}$
    \ENDFOR
    \STATE Form the sector-level predictor vector
    \[
        z_{i,t} = \big(f_{i,t}^{(1)}, \dots, f_{i,t}^{(K)}\big) \in \mathbb{R}^K
    \]
\ENDFOR
\RETURN $\{z_{i,t}\}$
\end{algorithmic}
\caption{Sector-level characteristics construction via PPCA. 
Here $\tau$ is the number of rolling windows, 
$N$ the number of sectors, 
$K$ the number of firm-level characteristics, 
$n_t$ the set of time points in the $t$-th rolling window (with cardinality $|n_t|$), 
and $n_i$ the number of firms in sector $i$. 
At each $(i,t,k)$, PPCA is applied to the $|n_t|\times n_i$ matrix $X_{i,t}^{(k)}$ of characteristic-$k$ values across firms and time. 
The first principal component $f_{i,t}^{(k)}$ is extracted, and the resulting sector-level predictor vector is 
$z_{i,t} = (f_{i,t}^{(1)}, \dots, f_{i,t}^{(K)})$.}
\end{algorithm}

We construct an ensemble of $16$ predictive models to forecast sector-level excess returns $r_{i,t+1}$ based on sector-level features $z_{i,t} \in \mathbb{R}^K$. The model class includes three linear benchmarks: Ordinary Least Squares (OLS), Principal Component Regression (PCR), and Least Absolute Shrinkage and Selection Operator (LASSO). The OLS model assumes a linear conditional mean $g(z_{i,t}; \theta) = z_{i,t}^\top \theta$ and estimates $\hat{\theta}$ by minimizing the residual sum of squares. PCR projects $z_{i,t}$ onto a lower-dimensional space using principal component analysis before fitting a linear model. LASSO augments the OLS objective with an $\ell_1$-penalty. Namely, $\mathcal{L}(\theta; \lambda) = \frac{1}{T} \sum_{t=1}^{T} \left(r_{i,t+1} - z_{i,t}^\top \theta\right)^2 + \lambda \sum_{j=1}^{K} |\theta_j|,$
which induces sparsity in $\theta$ and improves robustness in high-dimensional settings.

The remaining $13$ models are nonlinear machine learning methods. We include two tree-based algorithms: Random Forests (RF), which average predictions over an ensemble of decorrelated decision trees trained on bootstrap samples, and Gradient-Boosted Regression Trees (GBRT), which fit additive trees sequentially to minimize squared loss, 
$\hat{f}(z_{i,t}) = \sum_{b=1}^{B} \lambda \hat{f}^{(b)}(z_{i,t}),$
where $\lambda > 0$ is a shrinkage parameter. The remaining models are feedforward neural networks with increasing depth, denoted NN1 through NN12, where NN$k$ includes $k$ hidden layers with ReLU activation functions: $\mathrm{ReLU}(x) = \max(0, x).$

The final prediction is computed as $g(z_{i,t}; w) = x^{(L-1)\top} w^{(L-1)}$, where $x^{(L-1)}$ is the output of the last hidden layer and $w^{(L-1)}$ are the corresponding weights.

To evaluate out-of-sample predictive performance, we compute the $R^2_{oos}$ statistic for each sector $i$ over $\tau$ rolling windows:
\begin{equation}\label{eq:sectorR2}
R_{oos}^{2}(\tau)= 1-{\frac{\sum_{t=1}^\tau(r_{i,t}-\widehat{r}_{i,t})^{2}}{\sum_{t=1}^\tau r_{i,t}^{2}}},
\end{equation}
where $\widehat{r}_{i,t}$ is the model's forecast. Following \cite{guxiu:20}, the denominator omits demeaning to avoid using noisy sample means.

Since we forecast sector returns rather than individual stocks, our models facilitate sector-level capital allocation rather than asset-level optimization. For portfolio construction, we select top-performing sectors based on forecasts and assign weights to constituent stocks consistent with \eqref{eq:def_sector_returns}. This approach preserves the sector portfolio's structure and enables a feasible strategy that balances diversification with predictive signal strength. These portfolio returns are used in the evaluation of the sector rotation strategy in Section~\ref{sec:casestudy}.

\section{Performance Driven Ensemble of Models}\label{sec:ensemble}
The predictive accuracy of machine learning models exhibits substantial variation across sectors and over time. A model that underperforms in one sector during a particular period may outperform others in a different context or regime. This heterogeneity motivates the use of ensemble methods to aggregate forecasts. The most straightforward approach is the equally-weighted average, which assigns uniform weights to all models regardless of their historical performance. We adopt this naïve ensemble as a benchmark in our empirical analysis.

Let $\widehat{r}^{(l)}_{i,t}$ denote the forecast for sector $i$ at time $t$ from model $l$, for $l = 1,\ldots,L$. The ensemble prediction is defined as $\widetilde{r}_{i,t} = \left(\widehat{r}^{(1)}_{i,t},\ldots,\widehat{r}^{(L)}_{i,t}\right)^{\top} \mathbf{p}^{(t)},$
where $\mathbf{p}^{(t)} \in \mathbb{R}_+^L$ is a weight vector satisfying $\mathbbm{1}_L^\top \mathbf{p}^{(t)} = 1$. A more informed alternative is to compute optimal fixed weights \emph{offline} by solving a constrained regression that maximizes out-of-sample $R^2$:
\begin{equation}\label{eq:opt_p_problem}
  \widehat{\mathbf p}^*
  =\arg\max_{\substack{\mathbf p\in\mathbb{R}_+^L\\\mathbbm1_L^\top\mathbf p=1}}
    \left\{ 1 - 
      \frac{\tfrac1\tau\sum_{t=1}^\tau (r_t - \widehat{\mathbf r}_t^\top\mathbf p)^2}
           {\tfrac1\tau\sum_{t=1}^\tau r_t^2}
    \right\}.
\end{equation}
This quadratic program has a closed-form solution, analogous to stacked regression \cite{Bre:96}.

\begin{lemma}\label{lem:optimal_P_oosR2}
Under standard linear regression assumptions, the solution to \eqref{eq:opt_p_problem} is
\begin{equation}\label{eq:opt_p_solution}
\widehat{\mathbf p}^*
= \widehat{\mathbf{p}}_{\mathrm{OLS}} - (\widehat{\mathbf{R}}^\top \widehat{\mathbf{R}})^{-1} \mathbbm{1}_L \cdot
\frac{\mathbbm{1}_L^\top \widehat{\mathbf{p}}_{\mathrm{OLS}} - 1}{\mathbbm{1}_L^\top (\widehat{\mathbf{R}}^\top \widehat{\mathbf{R}})^{-1} \mathbbm{1}_L},
\end{equation}
where $\widehat{\mathbf{p}}_{\mathrm{OLS}} = (\widehat{\mathbf{R}}^\top \widehat{\mathbf{R}})^{-1} \widehat{\mathbf{R}}^\top \mathbf{r}$, $\widehat{\mathbf{R}} = [\widehat{\mathbf{r}}_1^\top, \ldots, \widehat{\mathbf{r}}_\tau^\top]^\top$, and $\mathbf{r} = [r_1, \ldots, r_\tau]^\top$.
\end{lemma}
\begin{proof}
The objective function in \eqref{eq:opt_p_problem} is quadratic and the equality constraints are linear. Hence, it is a convex problem with a unique minimizer. To determine it, we write the Lagrangian:
\[
\mathcal{L}(\mathbf{p}, \lambda) 
= \frac{1}{\tau}\sum_{t=1}^\tau \big(r_t - \widehat{\mathbf{r}}_t^{\top}\mathbf{p}\big)^2 
- 2\lambda \big(\mathbbm{1}_L^\top \mathbf{p} - 1\big),
\]
where $\widehat{\mathbf{r}}_t \in \mathbb{R}^L$ denotes the vector of $L$ model forecasts at time $t$.

Taking first-order conditions and writing in matrix form gives the system of $L+1$ equations:
\begin{equation}
\begin{bmatrix}
\widehat{\mathbf{R}}^{\top}\widehat{\mathbf{R}} & \mathbbm{1}_L \\
\mathbbm{1}_L^\top & 0
\end{bmatrix}
\begin{bmatrix}
\widehat{\mathbf{p}}^* \\
\lambda
\end{bmatrix}
=
\begin{bmatrix}
\widehat{\mathbf{R}}^\top \mathbf{r} \\
1
\end{bmatrix}, \nonumber
\end{equation}
where $\widehat{\mathbf{R}} = [\widehat{\mathbf{r}}_1^\top, \ldots, \widehat{\mathbf{r}}_\tau^\top]^\top \in \mathbb{R}^{\tau \times L}$ is the matrix of predicted returns, and $\mathbf{r} = [r_1,\ldots,r_\tau]^\top \in \mathbb{R}^\tau$ is the vector of realized returns.

Assuming invertibility of $\widehat{\mathbf{R}}^{\top}\widehat{\mathbf{R}}$, we obtain
\[
\widehat{\mathbf{p}}^* = \widehat{\mathbf{p}}_{\mathrm{OLS}} 
- (\widehat{\mathbf{R}}^{\top}\widehat{\mathbf{R}})^{-1}\mathbbm{1}_L \lambda,
\]
where $\widehat{\mathbf{p}}_{\mathrm{OLS}} = (\widehat{\mathbf{R}}^\top \widehat{\mathbf{R}})^{-1} \widehat{\mathbf{R}}^\top \mathbf{r}$ is the unconstrained OLS solution.
From the constraint equation we obtain\\ $\mathbbm{1}_L^\top(\widehat{\mathbf{R}}^{\top}\widehat{\mathbf{R}})^{-1}\mathbbm{1}_L \, \lambda 
= \mathbbm{1}_L^\top \widehat{\mathbf{p}}_{\mathrm{OLS}} - 1,$ which solves for $\lambda$. Substituting back yields the closed-form expression in Lemma~\ref{lem:optimal_P_oosR2}.
\end{proof}

Although \eqref{eq:opt_p_solution} is efficient to compute, it has critical limitations for forecasting. It produces static weights based on historical performance, assuming that model relevance remains constant. This assumption is often violated in financial environments where structural changes and regime shifts are frequent.

Moreover, the offline method does not address exploration exploitation. It emphasizes past accuracy but fails to consider whether low-weight models contribute redundant or complementary signals. As a result, it may overweight historically dominant models while underutilizing diverse ones.

To address these issues, we propose an online ensemble method based on the Multiplicative Weights Update Method (MWUM). MWUM updates weights sequentially using only recent forecast errors. It adapts to changing environments and promotes diversity by rewarding models that add non-redundant information. This makes it particularly effective for sequential forecasting in nonstationary settings.
\begin{figure}[t]
    \centering
    \includegraphics[scale=0.40]{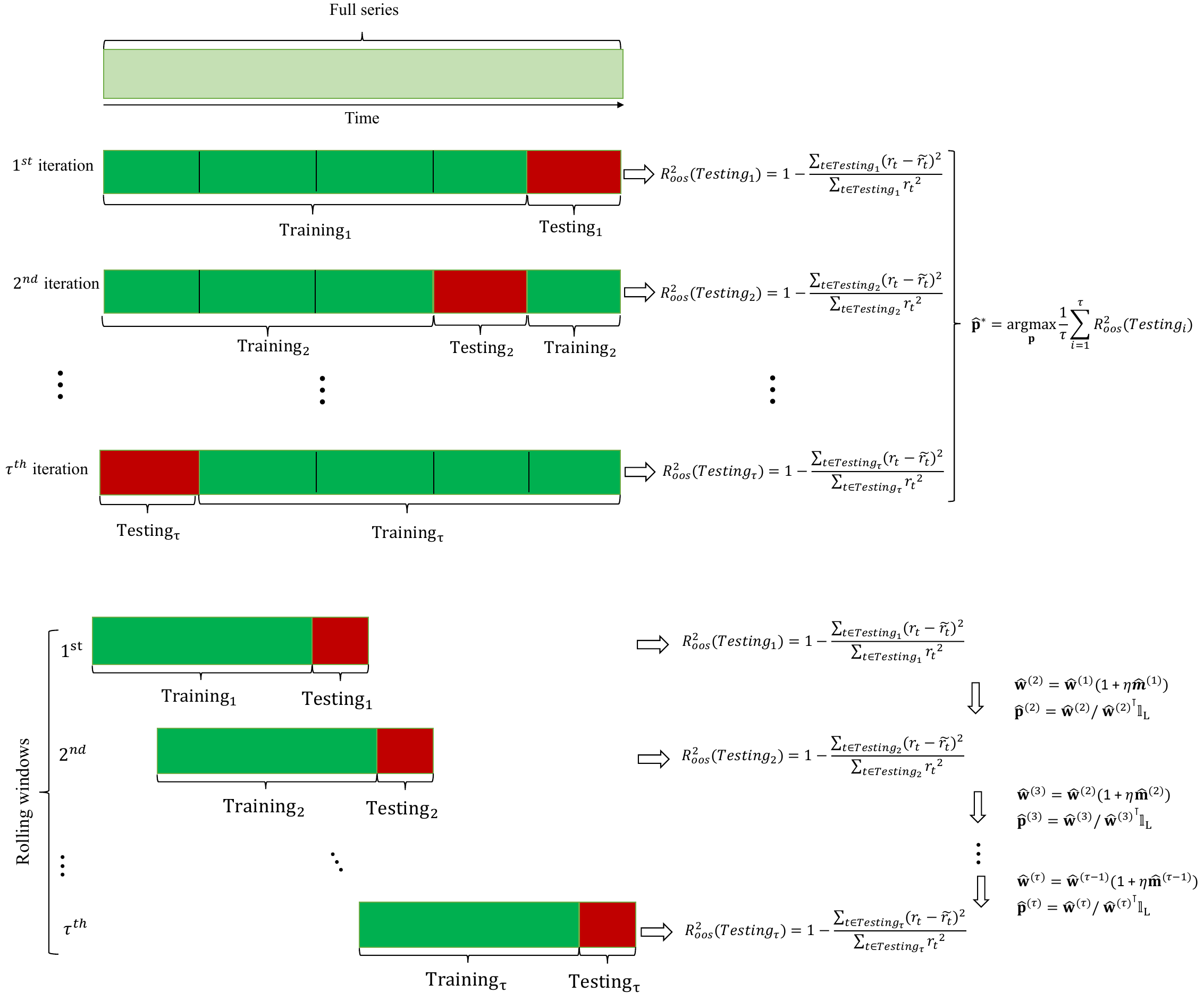}
\caption{Comparison of standard CV (top) and MWUM (bottom). CV relies on future data and retrains models at each step using gradient- or grid-based optimization, whereas MWUM updates weights online based on recent performance, without retraining or forward-looking data.}
    \label{fig:CV_vs_MWUM}
\end{figure}

Figure~\ref{fig:CV_vs_MWUM} contrasts MWUM with standard cross-validation. CV requires future observations, retraining, and meta-parameter search, which becomes impractical in high dimensions. MWUM avoids these issues by updating weights sequentially in real time, using fixed rules and past forecast errors.

\begin{algorithm}[h!]
\renewcommand{\thealgorithm}{}
\begin{algorithmic}[2]
\REQUIRE $L$ Models predicted returns $\{\widehat{\mathbf{r}}_t\}_{t=1}^\tau$; learning rate $\eta\in\mathbb{R}_+$
\ENSURE Ensemble forecasts $\{\widetilde{r}_t\}_{t=1}^\tau$ with weights $\mathbf{p}^{(t)}$
\STATE Initialize weights $\mathbf{w}^{(1)} = [1,\ldots,1]$
\FOR{$t = 2$ to $\tau$}
\STATE $\mathbf{p}^{(t)} = \mathbf{w}^{(t)} / \sum_{l=1}^L w_l^{(t)}$
\STATE Observe gains $\mathbf{m}^{(t)}$ (defined below)
\STATE Update: $\mathbf{w}^{(t+1)} = \mathbf{w}^{(t)}(1 + \eta \mathbf{m}^{(t)})$
\ENDFOR
\end{algorithmic}
\caption{Meta Multiplicative Weights Update Algorithm}
\label{al2}
\end{algorithm}

The gain function $\mathbf{m}^{(t)}(\mathbf{p}^{(t)})$ is defined as:
\begin{equation}\label{eq:costFunction}
\mathbf{m}^{(t)}(\mathbf{p}^{(t)}) = 
\underbrace{\left(\mathbbm{1}_L - \frac{(r_t \mathbbm{1}_L - \widehat{\mathbf{r}}_t)^2}{\widehat{\sigma}_t^2}\right)}_{\text{exploitation}}
\underbrace{- \frac{\widetilde{R}_t \mathbbm{1}_L}{\widehat{\sigma}_t^2}}_{\text{exploration}},
\end{equation}
where $r_t$ is the realized return, $\widehat{\sigma}_t^2$ is a consistent estimator of the returns variance at time $t$, and $\widetilde{R}_t = [\xi_{i,j,t}]$ is an $L \times L$ matrix with
\[
\xi_{i,j,t} =
\begin{cases}
\widehat{r}_{t,i} \widehat{r}_{t,j} p_j^{(t)} & \text{if } i \ne j \\
- \widehat{r}_{t,i}^2 (1 - p_i^{(t)}) & \text{otherwise}.
\end{cases}
\]
The proposed gain function $\mathbf{m}^{(t)}(\mathbf{p}^{(t)})$ naturally decomposes into two components that govern the trade-off between \textit{exploitation} and \textit{exploration} in the ensemble. 

\textit{Exploitation.} 
The first term captures each model's rolling-window squared forecast error relative to the estimated variance of returns, denoted by $\widehat{\sigma}_t^2$. 
In the empirical analysis, $\widehat{\sigma}_t^2$ is obtained from a bootstrap procedure applied to returns within the current rolling window. 
If model~$i$ achieves a forecast error smaller than $\widehat{\sigma}_t^2$, its gain contribution is positive, rewarding accuracy beyond one standard deviation of return fluctuations. 
When $\widehat{\sigma}_t^2$ is computed from raw squared returns without demeaning, this aligns with the na\"\i ve benchmark of zero mean returns used in the denominator of $R^2_{oos}$. 
Conversely, forecast errors exceeding $\widehat{\sigma}_t^2$ yield negative gains, penalizing poor performance. 
Thus, the exploitation term dynamically adjusts model weights according to predictive accuracy, systematically favoring models that consistently outperform the volatility benchmark.

\textit{Exploration.} The second term promotes diversity in the ensemble by comparing each model’s prediction with the current weighted forecast $\widetilde r_t = \widehat{\mathbf r}_t^\top \mathbf p^{(t)}$. Its behavior depends on the model’s current weight:  
(i) If $p_i^{(t)} \approx 1$, then $\widetilde r_t \approx \widehat r_{t,i}$ and the exploration contribution vanishes, meaning dominant models are not further adjusted.  
(ii) If $p_i^{(t)} \approx 0$ and the model predicts in the same direction as a dominant one ($\widehat r_{t,i}\widehat r_{t,j}>0$ for some $j$ with $p_j^{(t)}\approx 1$), it is penalized for redundancy.  
(iii) If $p_i^{(t)} \approx 0$ and the model predicts in the opposite direction ($\widehat r_{t,i}\widehat r_{t,j}<0$ for some $j$ with $p_j^{(t)}\approx 1$), then its weight is increased—provided the exploitation term indicates small error—thus rewarding complementary information. 

In this way, $\mathbf{m}^{(t)}(\mathbf{p}^{(t)})$ favors models that are both accurate and non-redundant, balancing exploitation with exploration. This mechanism implements an optimal bias–variance–diversity trade-off, as discussed in \cite{Wood:23}.

The following result provides the theoretical justification for our gain function. 
It shows that the average gain produced by the online ensemble converges to the out-of-sample $R^2$ performance metric, 
thereby validating $\mathbf{m}^{(t)}(\mathbf{p}^{(t)})$ as a principled surrogate objective for optimization.

\begin{assumption}[Moments, consistency, and bounded gain]\label{ass:basic}
We impose the following mild regularity conditions: (A0) \textit{Finite second moment and ergodicity.}  
The return process $(r_t)_{t\ge 1}$ is strictly stationary and ergodic with $\mathbb{E}[r_t^2]<\infty$.  
Consequently, $\frac{1}{\tau}\sum_{s=1}^{\tau} r_s^2 \xrightarrow{\mathrm{a.s.}} \mathbb{E}[r_1^2]
\quad \text{as } \tau\to\infty.$ (A1) \textit{Consistency of the moment estimator.} 
The estimator $\widehat\sigma_t^2$ of the second moment is almost surely consistent, that is, $\widehat\sigma_t^2 \xrightarrow{\mathrm{a.s.}} \mathbb{E}[r_1^2].$ (A2) \textit{Bounded gain.}  
There exists a constant $C<\infty$ such that for all $t$ and all weight vectors $\mathbf p$ in the probability simplex $\Delta_L$,
$\bigl|\,1-\mathbf m^{(t)}(\mathbf p)^{\!\top}\mathbf p\,\bigr|\le C,$
where $\mathbf m^{(t)}(\cdot)$ denotes the gain vector defined in~\eqref{eq:costFunction}.
\end{assumption}

\begin{lemma}[Asymptotic equivalence of average gain and out-of-sample performance]\label{lem:CostFunctionAsConvergenceToR2}
Let $\{\mathbf p^{(t)}\}_{t=1}^\tau\subset\Delta_L$ be any sequence of model weights, and let 
$\widehat{\mathbf r}_t=(\widehat r_{t,1},\ldots,\widehat r_{t,L})^\top$ denote the $L$ model predictions at time $t$.  
Define the ensemble forecast as $\widetilde r_t = \widehat{\mathbf r}_t^{\top}\mathbf p^{(t)}$, 
and the corresponding out-of-sample coefficient of determination as
\[
R^2_{oos}(\tau)
\;=\;
1-\frac{\sum_{t=1}^{\tau}(r_t-\widetilde r_t)^2}{\sum_{t=1}^{\tau} r_t^2}.
\]
For each $t$, let $\widehat\sigma_t^2$ denote an estimator of $\mathbb{E}[r_1^2]$.  
Then, for $\mathbf m^{(t)}\!\bigl(\mathbf p^{(t)}\bigr)$ defined in~\eqref{eq:costFunction}, 
and under Assumption~\ref{ass:basic},
\[
\left|\,R^2_{oos}(\tau)
-\frac{1}{\tau}\sum_{t=1}^{\tau}
\mathbf m^{(t)}\!\bigl(\mathbf p^{(t)}\bigr)^{\!\top}\mathbf p^{(t)}\,\right|
\xrightarrow{\mathrm{a.s.}} 0
\qquad \text{as } \tau\to\infty.
\]
\end{lemma}

\begin{proof}
We start by writing the $R^2_{oos}$ over $\tau$ periods as
\begin{align}
R^2_{oos}(\tau)
&= \frac{1}{\tau}\sum_{t=1}^{\tau}\!\left(1 - \frac{(r_t - \tilde r_t)^2}{\frac{1}{\tau}\sum_{s=1}^{\tau} r_s^2}\right),
\label{eq:R2-expanded}
\end{align}
\noindent\textbf{Step 1. Expanding the squared error:} $(r_t - \tilde r_t)^2
  = r_t^2 - 2r_t\,\tilde r_t + \tilde r_t^2
  = r_t^2 - 2r_t\,\widehat{\mathbf r}_t^\top \mathbf p^{(t)} + (\widehat{\mathbf r}_t^\top \mathbf p^{(t)})^2.$
The final term $(\widehat{\mathbf r}_t^\top \mathbf p^{(t)})^2$ is nonlinear in~$\mathbf p^{(t)}$.
To make its structure explicit, we decompose it into diagonal and cross products:
\begin{equation}\label{eq:square-decompose}
(\widehat{\mathbf r}_t^\top \mathbf p^{(t)})^2
= (\widehat{\mathbf r}_t^2)^\top (\mathbf p^{(t)})^2 + R_t,
\qquad
R_t := \sum_{i\neq j}\widehat r_{t,i}\widehat r_{t,j}\,p^{(t)}_i p^{(t)}_j,
\end{equation}
where $\widehat{\mathbf r}_t^2 = (\widehat r_{t,1}^2,\ldots,\widehat r_{t,L}^2)^\top$ and the notation $(\mathbf p^{(t)})^2$
denotes elementwise squaring. \\
\noindent\textbf{Step 2. Constructing a linear surrogate.}
The term $(r_t - \tilde r_t)^2/\widehat\sigma_t^2$ is nonlinear in~$\mathbf p^{(t)}$, which complicates the analysis.
To obtain a linear form, consider instead the weighted average, since $\mathbbm{1}_L^\top \mathbf p^{(t)} = 1$,
\begin{equation}\label{eq:linear-surrogate}
\left[\frac{(r_t\mathbbm{1}_L - \widehat{\mathbf r}_t)^2}{\widehat\sigma_t^2}\right]^{\!\top}\mathbf p^{(t)} = \frac{r_t^2 - 2r_t\,\widehat{\mathbf r}_t^\top\mathbf p^{(t)}
   + \widehat{\mathbf r}_t^{2\top}\mathbf p^{(t)}}{\widehat\sigma_t^2},
\end{equation}
which is \emph{linear} in $\mathbf p^{(t)}$.

\noindent\textbf{Step 3. Relating the nonlinear and linear terms via a Taylor expansion.}
Define the auxiliary bilinear function
\[
\varphi_t(\mathbf x, y)
 := \left(\alpha_t + \mathbf b_t^\top \mathbf x + \frac{R_t}{\widehat\sigma_t^2}\right)y,\]
 
with
$\alpha_t = \left(\frac{r_t^2\mathbbm{1}_L - 2r_t\,\widehat{\mathbf r}_t}{\widehat\sigma_t^2}\right)^\top \mathbf p^{(t)},$ and $\mathbf b_t = \frac{\widehat{\mathbf r}_t^2 \odot \mathbf p^{(t)}}{\widehat\sigma_t^2}.$ The notation $\odot$ denotes the Hadamard (elementwise) product.
The function $\varphi_t$ is affine in each argument and bilinear overall, which allows an exact first-order Taylor expansion around
$(\mathbf x_0, y_0) = (\mathbbm{1}_L, 1)$:
\begin{align}
\varphi_t(\mathbf p^{(t)}, y_t)
&= \varphi_t(\mathbf x_0, y_0)
 + \nabla_{\!\mathbf x}\varphi_t^\top(\mathbf p^{(t)} - \mathbbm{1}_L)
 + \partial_y\varphi_t (y_t - 1)\nonumber \\
 &+ \nabla_{\!\mathbf x}\partial_y\varphi_t^\top(\mathbf p^{(t)} - \mathbbm{1}_L)(y_t - 1),
\label{eq:taylor}
\end{align}
where $y_t := \widehat\sigma_t^2 / \bigl(\frac{1}{\tau}\sum_{s=1}^{\tau}r_s^2\bigr)$. Evaluating $\varphi_t$ at the two points of interest gives:
\[
\varphi_t(\mathbf p^{(t)}, y_t)
 = \frac{(r_t - \tilde r_t)^2}{\frac{1}{\tau}\sum_{s=1}^{\tau}r_s^2},
\qquad
\varphi_t(\mathbf x_0, y_0)
 = \left[\frac{(r_t\mathbbm{1}_L - \widehat{\mathbf r}_t)^2}{\widehat\sigma_t^2}\right]^{\!\top}\mathbf p^{(t)} + \frac{R_t}{\widehat\sigma_t^2}.
\]
Substituting these into \eqref{eq:taylor} and grouping terms yields
\begin{equation}\label{eq:ratio-decomposition}
\frac{(r_t - \tilde r_t)^2}{\frac{1}{\tau}\sum_{s=1}^{\tau}r_s^2}
= \left[\frac{(r_t\mathbbm{1}_L - \widehat{\mathbf r}_t)^2}{\widehat\sigma_t^2}\right]^{\!\top}\mathbf p^{(t)}
 + \frac{R_t}{\widehat\sigma_t^2} + \varepsilon_t,
\end{equation}
where $\varepsilon_t$ collects all first-order remainder terms from $(y_t - 1)$ and $(\mathbf p^{(t)} - \mathbbm{1}_L)$. Substituting \eqref{eq:ratio-decomposition} into the expression for $R^2_{oos}(\tau)$ in~\eqref{eq:R2-expanded} gives
\[
R^2_{oos}(\tau)
= \frac{1}{\tau}\sum_{t=1}^{\tau}\!\left(
  1 - \left[\frac{(r_t\mathbbm{1}_L - \widehat{\mathbf r}_t)^2}{\widehat\sigma_t^2}\right]^{\!\top}\mathbf p^{(t)}
  - \frac{R_t}{\widehat\sigma_t^2} - \varepsilon_t
\right).
\]
Recognizing the definition of the gain function $\mathbf m^{(t)}(\mathbf p^{(t)})$ from~\eqref{eq:costFunction},
this can be rewritten compactly as
\begin{equation}\label{eq:R2-diff}
R^2_{oos}(\tau)
 = \frac{1}{\tau}\sum_{t=1}^{\tau}\mathbf m^{(t)}(\mathbf p^{(t)})^{\!\top}\mathbf p^{(t)} + \varepsilon_\tau,
\end{equation}
where the aggregated remainder is
\[
\varepsilon_\tau := -\frac{1}{\tau}\sum_{t=1}^{\tau}\!\Big(1 - \mathbf m^{(t)}(\mathbf p^{(t)})^{\!\top}\mathbf p^{(t)}\Big)(y_t - 1).
\]
\noindent\textbf{Step 4. Convergence of the remainder term.}
By Assumptions~(A0)–(A1),
both $\widehat\sigma_t^2$ and $\frac{1}{\tau}\sum_{s=1}^{\tau}r_s^2$ are consistent estimators of $\mathbb E[r_1^2]$.
Hence, by the continuous-mapping theorem, $y_t = \frac{\widehat\sigma_t^2}{\frac{1}{\tau}\sum_{s=1}^{\tau}r_s^2}
\xrightarrow{\text{a.s.}} 1.$ Under Assumption~(A2), $\big|1 - \mathbf m^{(t)}(\mathbf p^{(t)})^{\!\top}\mathbf p^{(t)}\big|\le C$ for some constant~$C<\infty$.
Therefore,
\[
\big|\varepsilon_\tau\big|
\le \frac{1}{\tau}\sum_{t=1}^{\tau}\!
 \big|1 - \mathbf m^{(t)}(\mathbf p^{(t)})^{\!\top}\mathbf p^{(t)}\big|\,|y_t - 1|
\le C\,\frac{1}{\tau}\sum_{t=1}^{\tau}|y_t - 1|.
\]
Because $y_t\to 1$ almost surely, Cesàro’s lemma implies
$\frac{1}{\tau}\sum_{t=1}^{\tau}|y_t - 1| \to 0$ almost surely, and hence $\varepsilon_\tau \to 0$ a.s. Combining the results above, from \eqref{eq:R2-diff} we obtain
\[
\left|R^2_{oos}(\tau)
  - \frac{1}{\tau}\sum_{t=1}^{\tau}\mathbf m^{(t)}(\mathbf p^{(t)})^{\!\top}\mathbf p^{(t)}\right|
 = |\varepsilon_\tau|
 \xrightarrow{\text{a.s.}} 0.
\]
\end{proof}
The following lemma provides a high-probability bound (non-asymptotic, for finite $\tau$) showing that the average gain concentrates around the out-of-sample $R^2_{oos}(\tau)$.

\begin{lemma}\label{lem:CostFunctionAsConvergenceToR2Concentration}
Assume $r_t \overset{\mathrm{iid}}{\sim} \mathcal{N}(0, \sigma_r^2)$. Then for any weight sequence $\{\mathbf p^{(t)}\}_{t=1}^\tau$ and any $\varepsilon > 0$,
\[
\mathbb{P}\left( 
  \left| R^2_{oos}(\tau) - \frac{1}{\tau} \sum_{t=1}^\tau \mathbf{m}^{(t)}(\mathbf{p}^{(t)})^\top \mathbf{p}^{(t)} \right| > \varepsilon 
\right) \leq \frac{2(1 + L)}{\varepsilon\, \sigma_r\, \tau^2 \sqrt{\tfrac{8}{\tau^2} - \tfrac{6}{\tau} + 1}}.
\]
\end{lemma}

\begin{proof}
We apply Chebyshev’s inequality to the residual term from the last proof to get
\[
\mathbb{P}\left( 
  \left| \frac{1}{\tau} \sum_{t=1}^\tau (1 - \mathbf{m}^{(t)}(\mathbf{p}^{(t)})^\top \mathbf{p}^{(t)})(y_t - y_0) \right| > \varepsilon 
\right) \leq \frac{\mathbb{E}[|Z|]}{\varepsilon},
\]
where $Z := \frac{1}{\tau} \sum_t (1 - \mathbf{m}^{(t)}(\mathbf{p}^{(t)})^\top \mathbf{p}^{(t)})(y_t - y_0)$.

Applying Young’s inequality for all $\lambda > 0$ gives:
\[
\mathbb{E}[|Z|] \leq \frac{\lambda^2}{2} \mathbb{E}\left\| \frac{1}{\sqrt{\tau}} (y_{(\cdot)} - \mathbf{1}_\tau y_0) \right\|_2^2 + \frac{1}{2\lambda^2} \mathbb{E} \left\| \frac{1}{\sqrt{\tau}} \left( \mathbf{1}_\tau - \mathbf{m}^{(\cdot)}(\mathbf{p}^{(\cdot)})^\top \mathbf{p}^{(\cdot)} \right) \right\|_2^2.
\]

The second term is bounded by: $\mathbb{E}\left\| \cdot \right\|_2^2 \leq (1 + L)^2.$ For the first term, since $y_t = \widehat{\sigma}_t^2 / \frac{1}{\tau} \sum_{t=1}^\tau r_t^2$ and $r_t \sim \mathcal{N}(0, \sigma_r^2)$, we use standard results for inverse-$\chi^2$ distributions to obtain:
\[
\mathbb{E} \left\| \frac{1}{\sqrt{\tau}} (y_{(\cdot)} - \mathbf{1}_\tau y_0) \right\|_2^2 
\leq \frac{4}{\sigma_r^2 \tau^2 (\tau - 2)(\tau - 4)}.
\]

Combining these, we obtain:
\begin{align}
&\mathbb{P}\left( 
  \left| R^2_{oos}(\tau) - \frac{1}{\tau} \sum_{t=1}^\tau \mathbf{m}^{(t)}(\mathbf{p}^{(t)})^\top \mathbf{p}^{(t)} \right| > \varepsilon 
\right) \nonumber \\
& \leq \frac{\lambda^2}{2} \cdot \frac{4}{\sigma_r^2 \tau^2 (\tau - 2)(\tau - 4)\varepsilon} + \frac{(1 + L)^2}{2\lambda^2 \varepsilon}.\label{eq:bound}
\end{align}
Optimizing \eqref{eq:bound} with respect to $\lambda$ yields the stated bound.
\end{proof}

This result shows that the probability of deviation between the empirical $R^2_{oos}$ and the gain-based approximation decays at the rate $\tau^{-2}$, confirming the concentration behavior of our ensemble's learning objective.

In the empirical analysis presented in Section~\ref{sec:casestudy}, we conduct a rolling-window backtest spanning 35 years. Given the structural instability of financial markets, it is unrealistic to assume that a fixed ensemble specification remains optimal throughout such an extended period. To address this, we adaptively determine the ensemble’s learning rate $\eta>0$ based on recent forecast performance. Specifically, in each rebalancing period, we select a \textit{feasible} learning rate $\eta^*$ by choosing the value that achieved the best out-of-sample performance over the prior 12 months from a predefined grid of candidates. As shown in Section~\ref{sec:casestudy}, this adaptive online ensemble consistently outperforms individual models, the equally-weighted average ensemble, and an offline ensemble optimized using \eqref{eq:opt_p_solution} in each rolling-window.

\section{Empirical Study}\label{sec:casestudy}

We conduct our empirical analysis using monthly stock return data from the Center for Research in Security Prices (CRSP), covering the period from March 1957 to December 2021. Following \cite{guxiu:20}, we use 94 firm-level predictive characteristics spanning categories such as past returns, profitability, valuation, and trading frictions. Of these, 61 are updated annually, 13 quarterly, and 20 monthly. To prevent look-ahead bias, we apply standard publication lags: one month for monthly variables, four months for quarterly variables, and six months for annual variables. Thus, returns from $t$ to $t+1$ are forecasted using the most recently available data at time $t$, i.e., characteristics disclosed at $t-1$, $t-4$, and $t-6$ respectively. Sectors are defined based on the first two digits of the Standard Industrial Classification (SIC) codes.

The dataset is divided into a 30-year training window (1957–1986) and a 35-year out-of-sample test period (1987–2021). To reduce computational overhead, models are refitted annually rather than monthly. At the start of each year, the training set is expanded by one year, and monthly forecasts are generated for the following 12 months. Within each rolling training window (approximately 360 observations), we apply 5-fold cross-validation to tune model-specific hyperparameters.

To ensure robust ensemble performance, we exclude models that systematicaly underperform over time. Specifically, we omit OLS and shallow neural networks with fewer than six layers, which often yield negative out-of-sample $R^2$ values and are dominated by regularized methods such as LASSO and deeper network architectures.

We begin by evaluating the predictive performance of individual models across sectors. Table~\ref{tab:oos2} reports the average out-of-sample $R^2$ values, as defined in \eqref{eq:sectorR2}, for both equally- and capitalization-weighted sector returns based on \eqref{eq:def_sector_returns}. Across all models, performance is consistently higher under capitalization-weighted returns, reflecting the fact that small-cap stocks exhibit more idiosyncratic volatility, while our predictors capture systematic sector-level variation that is more relevant for large, liquid firms.

Among individual models, the LASSO exhibits the highest average out-of-sample $R^2_{\text{oos}}$, followed by the deeper neural network specifications. Table~\ref{tab:oos2} further reports results for three ensemble benchmarks. The \textit{Simple Average} ensemble assigns equal weights to all constituent models and serves as a naive baseline. The \textit{Offline Ensemble} implements the optimal static weighting scheme obtained from the fixed-weight solution in \eqref{eq:opt_p_solution}. The \textit{Exploitation Ensemble} updates weights dynamically but relies solely on the exploitation component of the gain function in \eqref{eq:costFunction}. Finally, the \textit{Online Ensemble} adaptively adjusts weights using the complete gain function and selects the learning rate $\eta^*$ within each rolling window based exclusively on past forecast performance over the preceding 12 months, ensuring full implementability without look-ahead bias.

Empirically, the Online Ensemble consistently outperforms all individual models and benchmark ensembles, achieving the highest $R^2_{\text{oos}}$ across sectors and evaluation periods. Its performance demonstrates the effectiveness of dynamically reweighting models according to their recent predictive accuracy and stability. Relative to the Exploitation Ensemble, the improvement underscores the importance of the exploration term in \eqref{eq:costFunction}, which promotes model diversity and mitigates overfitting to transient patterns. Accordingly, all subsequent analyses—including the sector rotation experiments—are conducted using forecasts generated by the Online Ensemble.

\begin{table*}[t]
  \centering
  \scriptsize
  \caption[Forecasting performance \(R^2_{oos}\)]{%
    Summary of forecasting performance results in terms of average
    across-sectors percentage \(R_{oos}^{2}\).  
    \textbf{Columns:} ten ML models; the simple-averaging ensemble; offline ensemble; Exploitation Ensemble;
    and our feasible online ensemble.  
    \textbf{Rows:} equally-weighted vs.\ cap-weighted sector returns.%
  }
  \label{tab:oos2}
  \resizebox{\textwidth}{!}{%
    \begin{tabular}{@{}l*{14}{c}@{}}
      \toprule
      & PCR & LASSO & GBRT & NN6 & NN7 & NN8 & NN9 & NN10
      & NN11 & NN12 
      & \begin{tabular}{c}Simple\\Average\end{tabular}
      & \begin{tabular}{c}Offline \\Ensemble\end{tabular}
      & \begin{tabular}{c}Exploitation \\Ensemble\end{tabular}
      & \begin{tabular}{c}Online \\Ensemble\end{tabular} \\
      \midrule
      Average return
        & -0.62 & 0.95 & 0.56 & -0.06 & 0.83 & 0.53 & 0.24 & 0.77 
        & 0.62 & 0.75 
        & 0.90 & 0.93&0.99 & \textbf{1.19}\\
      Weighted return
        &  0.38 & 1.11 & 0.89 &  0.35 & 0.36 & 1.06 & 0.89 & 0.95 
        & 0.80 & 0.75 
        & 1.07 & 1.08 &1.09& \textbf{1.12}\\
      \bottomrule
    \end{tabular}
  }
\end{table*}

Almost all models in Table~\ref{tab:oos2} produce positive $R^2_{oos}$ values, and the best-performing ensembles exceed the values reported in \cite{guxiu:20} for individual stock returns. This highlights the advantage of aggregating both characteristics and returns at the sector level, which improves signal-to-noise ratios and enhances predictability.

By aggregating forecasts across models, the online ensemble eliminates the need for model selection and enables more robust decision-making. As we will show in the next section, accounting for model heterogeneity across sectors further improves performance in a sector rotation strategy.

\begin{table}[H]
\tiny
\centering
\caption{Portfolio statistics (``Annual Return", ``Annual Volatility", ``Annual Sharpe", ``Max. Drawdown", and ``Annual Sortino") on portfolios. The predicted returns are calculated from meta-strategy. Each month we sort sectors in ascending order into quantile portfolios (``Bottom 5", ``41-55", ``21-40", ``6-20", and ``Top 5'') on the basis of predicted returns. The table also includes the benchmark portfolios market index (``S\&P 500") and (``1/N"), which is the average return across the 50 sectors. Panel I (A-B): 1987-2021. Panel II (A-B): Recent COVID-19 Period 2020-2021.}
\label{statistics_po}

\begin{tabular}{|lllllll|llll|}
\hline\noalign{\smallskip}
Statistics (1987-2021)&	\begin{tabular}{@{}c@{}}S\&P   \\ 500\end{tabular}	&	1/N 	&	Bottom 5	&	41-55	&	21-40	&	6-20	&	Top 5	&	\begin{tabular}{@{}c@{}}  \\ 5BPs\end{tabular}	&	\begin{tabular}{@{}c@{}}Net  \\ 10BPs\end{tabular}	&	\begin{tabular}{@{}c@{}}  \\ 15BPs\end{tabular} 	
\\
\noalign{\smallskip}\hline\noalign{\smallskip}
\begin{tabular}{@{}l@{}}Panel I.A.\\ \underline{Equally-Weighted Returns}\end{tabular}&&&&&&&&&&\\
Annual Return	&	0.0889	&	0.1048	&	0.0661	&	0.0921	&	0.0983	&	0.1228	&	0.1415	&	0.1368	&	0.1322	&	0.1275	\\
Annual Volatility	&	0.1504	&	0.1971	&	0.2304	&	0.2041	&	0.1951	&	0.1973	&	0.2153	&	0.2152	&	0.2151	&	0.2150	\\
Annual Sharpe	&	0.5912	&	0.5317	&	0.2869	&	0.4511	&	0.5037	&	0.6225	&	0.6570	&	0.6357	&	0.6143	&	0.5931	\\
Max. Drawdown	&	0.5256	&	0.6219	&	0.6639	&	0.6716	&	0.6204	&	0.5937	&	0.5233	&	0.5263	&	0.5292	&	0.5321	\\
Annual Sortino	&	0.9487	&	0.9114	&	0.5903	&	0.8105	&	0.8640	&	1.0452	&	1.1465	&	1.1122	&	1.0782	&	1.0443	\\
\begin{tabular}{@{}l@{}}Panel I.B.\\ \underline{Capital-Weighted Returns}\end{tabular}&&&&&&&&&&\\
Annual Return	&	0.0889	&	0.1094	&	0.0950	&	0.0930	&	0.1117	&	0.1126	&	0.1385	&	0.1342	&	0.1300	&	0.1258	\\
Annual Volatility	&	0.1504	&	0.1703	&	0.2064	&	0.1826	&	0.1686	&	0.1686	&	0.1883	&	0.1883	&	0.1883	&	0.1884	\\
Annual Sharpe	&	0.5912	&	0.6422	&	0.4602	&	0.5093	&	0.6625	&	0.6676	&	0.7355	&	0.7129	&	0.6904	&	0.6679	\\
Max. Drawdown	&	0.5256	&	0.5464	&	0.5995	&	0.5966	&	0.5617	&	0.4967	&	0.4825	&	0.4862	&	0.4899	&	0.4935	\\
Annual Sortino	&	0.9487	&	1.0245	&	0.8132	&	0.8327	&	1.0657	&	1.0693	&	1.2190	&	1.1832	&	1.1477	&	1.1124	\\
\noalign{\smallskip}\hline
\hline\noalign{\smallskip}
Statistics (2020-2021)&	\begin{tabular}{@{}c@{}}S\&P   \\ 500\end{tabular}	&	1/N 	&	Bottom 5	&	41-55	&	21-40	&	6-20	&	Top 5	&	\begin{tabular}{@{}c@{}}  \\ 5BPs\end{tabular}	&	\begin{tabular}{@{}c@{}}Net  \\ 10BPs\end{tabular}	&	\begin{tabular}{@{}c@{}}  \\ 15BPs\end{tabular} 
\\
\noalign{\smallskip}\hline\noalign{\smallskip}
\begin{tabular}{@{}l@{}}Panel II.A. \\ \underline{Equally-Weighted Returns}\end{tabular}&&&&&&&&&&\\
Annual Return	&	0.2162	&	0.2847	&	0.1915	&	0.3138	&	0.1925	&	0.3350	&	0.5040	&	0.4957	&	0.4875	&	0.4794	\\
Annual Volatility	&	0.1954	&	0.3147	&	0.3623	&	0.3300	&	0.3336	&	0.2837	&	0.3324	&	0.3322	&	0.3320	&	0.3319	\\
Annual Sharpe	&	1.1067	&	0.9049	&	0.5287	&	0.9507	&	0.5769	&	1.1807	&	1.5163	&	1.4922	&	1.4683	&	1.4444	\\
Max. Drawdown	&	0.2000	&	0.3524	&	0.4569	&	0.3568	&	0.3896	&	0.2750	&	0.3173	&	0.3181	&	0.3190	&	0.3198	\\
Annual Sortino	&	1.8400	&	1.5121	&	1.0014	&	1.6386	&	1.0437	&	2.0379	&	2.4006	&	2.3646	&	2.3287	&	2.2931	\\
\begin{tabular}{@{}l@{}}Panel II.B. \\ \underline{Capital-Weighted Returns}\end{tabular}&&&&&&&&&&\\
Annual Return	&	0.2162	&	0.2099	&	0.1955	&	0.1335	&	0.2396	&	0.2126	&	0.3073	&	0.3021	&	0.2970	&	0.2920	\\
Annual Volatility	&	0.1954	&	0.2457	&	0.3273	&	0.2792	&	0.2361	&	0.2240	&	0.2410	&	0.2410	&	0.2410	&	0.2409	\\
Annual Sharpe	&	1.1067	&	0.8545	&	0.5973	&	0.4782	&	1.0147	&	0.9495	&	1.2749	&	1.2538	&	1.2327	&	1.2117	\\
Max. Drawdown	&	0.2000	&	0.2996	&	0.4178	&	0.3489	&	0.2835	&	0.2581	&	0.2300	&	0.2313	&	0.2327	&	0.2340	\\
Annual Sortino	&	1.8400	&	1.3506	&	1.0135	&	0.8278	&	1.6604	&	1.4684	&	2.1066	&	2.0694	&	2.0325	&	1.9959	\\
\noalign{\smallskip}\hline
\end{tabular}

\end{table}

To evaluate the practical effectiveness of our ensemble forecasts, we implement a sector rotation strategy based on predicted returns from the feasible online ensemble. Each month, sectors are sorted in ascending order and grouped into five quantile portfolios: ``Bottom 5'', ``41--55'', ``21--40'', ``6--20'', and ``Top 5''. Stocks within each sector are held using the same weighting scheme as defined in \eqref{eq:def_sector_returns}, and we evaluate both equally-weighted and capitalization-weighted implementations.

As benchmarks, we include an equally-weighted portfolio across all sectors ($1/N$) and the S\&P 500 index. Table~\ref{statistics_po} reports annualized return, volatility, Sharpe ratio, maximum drawdown, and Sortino ratio for all quantile portfolios and benchmarks, covering the full out-of-sample period (1987–2021; Panel I) and the COVID-19 period (2020–2021; Panel II). We isolate the pandemic period due to heightened volatility and sector dispersion.

During the COVID-19 crash, the S\&P 500 declined by 34\% in February–March 2020, followed by a rapid but uneven recovery. While technology, healthcare, and consumer staples sectors rebounded quickly, others such as energy, industrials, and financials lagged. The ensemble forecasts captured these dynamics effectively, leading to substantial gains for the ``Top 5'' portfolio.

The strategy shows a clear monotonic relationship between predicted returns and realized performance. Annualized returns increase from 6.61\% for the ``Bottom 5'' to 14.15\% for the ``Top 5'' portfolios over the full sample, and from 19.15\% to 50.4\% during the pandemic (Panel II.A), using equally-weighted returns. The ``Top 5'' portfolio consistently achieves the highest Sharpe and Sortino ratios, while maintaining volatility and drawdowns comparable to or better than benchmarks. Similar results hold for capitalization-weighted returns (Panel II.B).

To account for trading frictions, we evaluate the ``Top 5'' strategy under transaction costs of 5, 10, and 15 basis points per percentage point of turnover. Even after costs, net returns remain strong, confirming the strategy’s robustness and implementability.

Over the full out-of-sample period, the ``Top 5'' strategy achieves Sortino ratios of 1.14 (equally-weighted) and 1.21 (capital-weighted). During the COVID-19 period, performance improves markedly, with Sortino ratios rising to 2.4 and 2.1, respectively. Sharpe ratios over this period also exceed both benchmarks and the historical Top 5 portfolios. These results demonstrate that our ensemble-guided strategy is able to dynamically select outperforming sectors, even during periods of severe market stress.

\begin{figure*}[t]
\centering
\begin{subfigure}{0.48\textwidth}
  \centering
  \includegraphics[width=\linewidth]{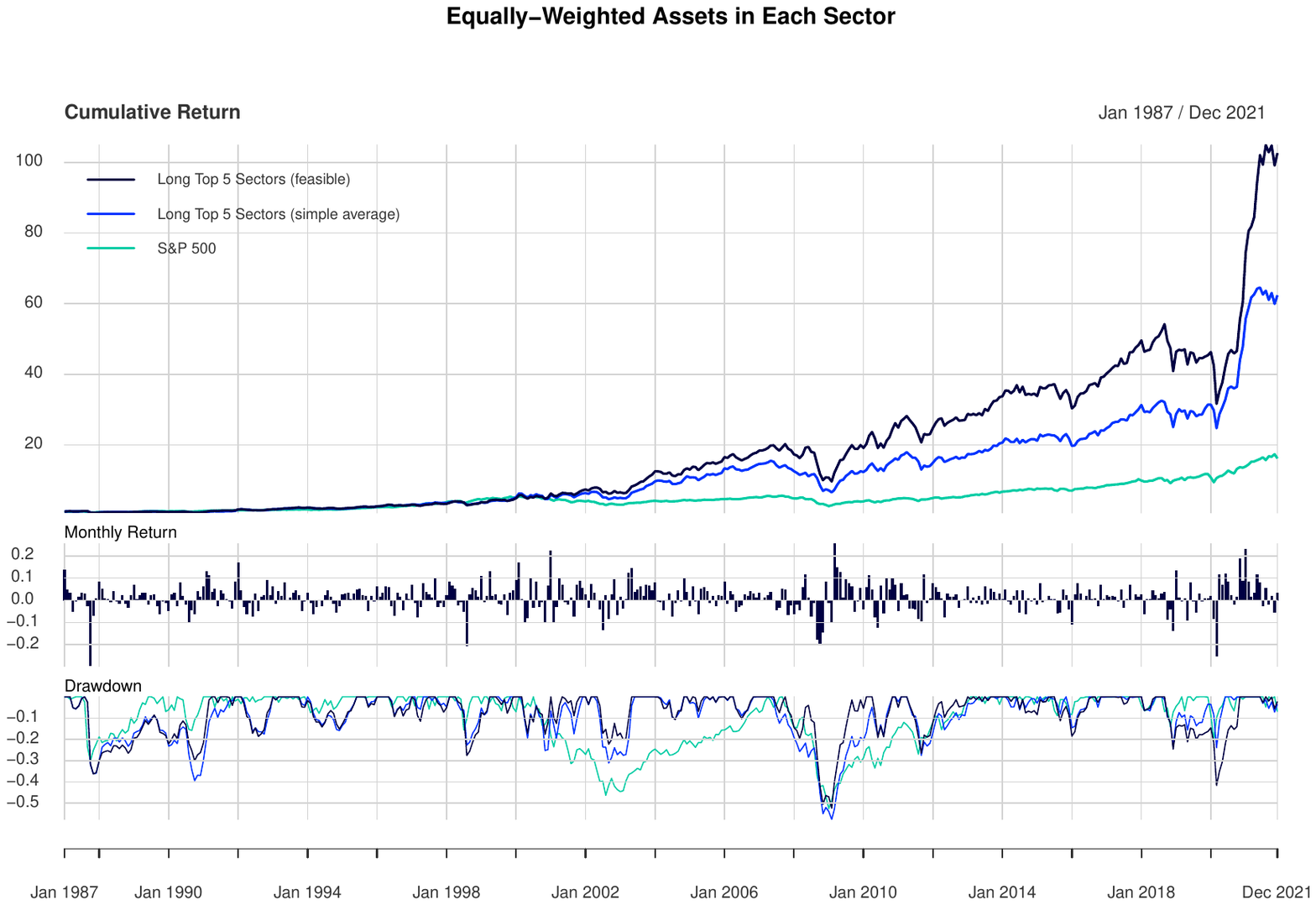}
  \caption{Equally-weighted returns}
  \label{performance:average}
\end{subfigure}
\hfill
\begin{subfigure}{0.48\textwidth}
  \centering
  \includegraphics[width=\linewidth]{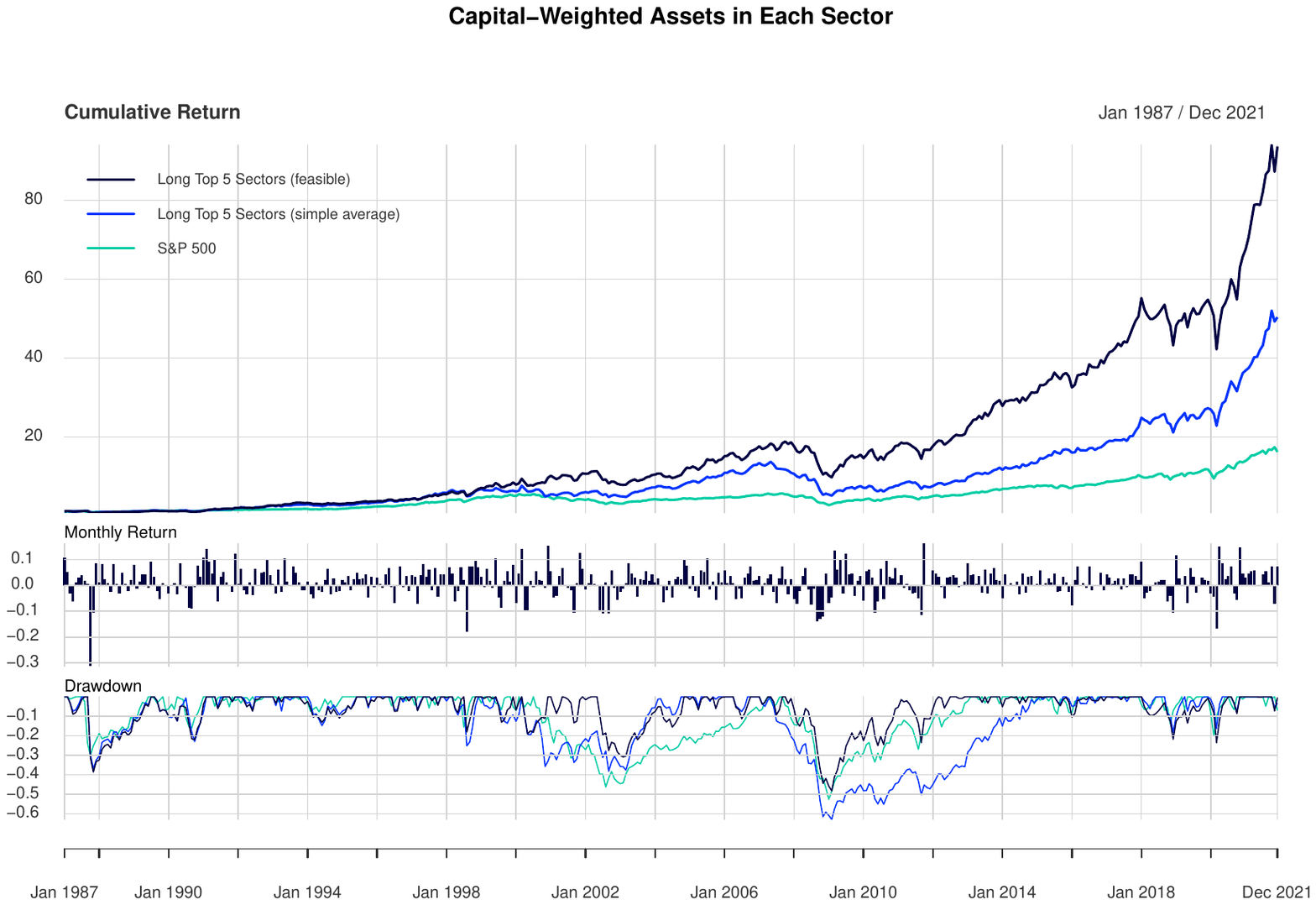}
  \caption{Capital-weighted returns}
  \label{performance:weighted}
\end{subfigure}
\caption{Performance of the ``Top 5'' sector strategy from Jan.\ 1987 to Dec.\ 2021, using equally-weighted (left) and capital-weighted (right) returns as defined in \eqref{eq:def_sector_returns}. \textbf{Top:} cumulative returns vs.\ S\&P 500. \textbf{Middle:} monthly returns. \textbf{Bottom:} drawdowns.}
\label{fig:performance_summary}
\end{figure*}

Although Table~\ref{statistics_po} reports aggregate performance metrics, it is essential to examine the evolution of strategy performance over time—particularly for machine learning methods, which can lose their edge as market participants adapt or as new information becomes widely available. To assess persistence, we analyze the monthly performance of the ``Top 5'' sector portfolio. Figures~\ref{performance:average} and \ref{performance:weighted} display cumulative returns, monthly returns, and drawdowns for the strategy versus the S\&P 500 index, using equally- and capitalization-weighted returns as defined in \eqref{eq:def_sector_returns}.
The results indicate that our strategy has not been arbitraged away. Its cumulative outperformance relative to the market grows steadily, especially around periods of structural shifts—most notably during the COVID-19 crisis. During this period, investors rapidly rotated into sectors such as technology, healthcare, and consumer staples. Our ensemble effectively captured this shift, enabling substantial gains for the ``Top 5'' strategy.

Both weighting schemes yield favorable drawdown behavior. The ``Top 5'' portfolios consistently exhibit smaller peak-to-trough declines and faster recoveries compared to the market during major downturns such as the dot-com crash (2002) and the global financial crisis (2008). This suggests that the strategy provides not only outperformance but also downside protection.

Finally, the robustness of the strategy across weighting schemes indicates that its success is not reliant on small-cap, illiquid stocks. Capitalization-weighted implementation enables exposure to more liquid names, improving tradeability and scalability—important considerations for institutional investors.

Table~\ref{excess_returns_on_portfolios} reports additional performance diagnostics for our feasible ensemble, including predicted returns, average excess returns, and alphas estimated from regressions on standard asset pricing models: (i) CAPM \cite{Sharpe:64}, (ii) Fama-French three-factor model \cite{FAMA:93}, and (iii) Carhart four-factor model \cite{Carhart:97}.

\begin{table*}[t]
\centering
\caption{\textbf{Returns on portfolios selected by the meta-strategy.} 
Average monthly returns from Jan.\ 1987 to Dec.\ 2021 on quantile portfolios sorted by predicted returns (``Bottom~5'' to ``Top~5''). 
``Top--Bottom'' shows return spreads. 
Top row reports average predicted returns; subsequent rows show excess returns and alphas from the CAPM~\cite{Sharpe:64}, 
Fama--French~\cite{FAMA:93}, and Carhart~\cite{Carhart:97} models. 
Panel~A reports equally weighted, Panel~B value weighted results. 
$t$-values (Newey--West~\cite{Newey:87}) are in parentheses. 
*, **, and *** denote significance at the 10\%, 5\%, and 1\% levels.}
\label{excess_returns_on_portfolios}

\resizebox{\textwidth}{!}{%
\begin{tabular}{lcccccc@{\hskip 1cm}cccccc}
\hline\noalign{\smallskip}
\textbf{Statistics}
& \multicolumn{6}{c}{\textbf{Panel A. Equally-Weighted Returns}} 
& \multicolumn{6}{c}{\textbf{Panel B. Capitalization-Weighted Returns}} \\[2pt]
\cmidrule(lr){2-7}\cmidrule(lr){8-13}
 & Bottom 5 & 41--55 & 21--40 & 6--20 & Top 5 & Top--Bottom 
 & Bottom 5 & 41--55 & 21--40 & 6--20 & Top 5 & Top--Bottom \\ 
\hline\noalign{\smallskip}

Predicted return 
& 0.0019 & 0.0067 & 0.0096 & 0.0125 & 0.0172 & 0.0177 
& 0.0008 & 0.0055 & 0.0084 & 0.0116 & 0.0178 & 0.0195 \\

Excess return 
& 0.0076** & 0.0091*** & 0.0095*** & 0.0114*** & 0.0130*** & 0.0079*** 
& 0.0094*** & 0.0089*** & 0.0101*** & 0.0101*** & 0.0124*** & 0.0054** \\

& (2.06) & (2.87) & (3.05) & (3.63) & (3.96) & (4.22) 
& (3.04) & (3.66) & (4.16) & (4.56) & (4.81) & (2.47) \\

CAPM alpha 
& 0.0014 & 0.0032 & 0.0036* & 0.0054*** & 0.0071*** & 0.0081*** 
& 0.0032 & 0.0027* & 0.0043*** & 0.0043*** & 0.0064*** & 0.0056** \\

& (0.53) & (1.40) & (1.85) & (2.77) & (3.03) & (4.13) 
& (1.32) & (1.85) & (3.25) & (4.68) & (3.71) & (2.53) \\

3F alpha 
& -0.0016 & 0.0004 & 0.0010 & 0.0028*** & 0.0047*** & 0.0088*** 
& 0.0003 & 0.0001 & 0.0019** & 0.0022*** & 0.0042*** & 0.0064*** \\

& (-0.99) & (0.27) & (0.93) & (2.62) & (2.97) & (4.59) 
& (0.20) & (0.13) & (2.37) & (2.79) & (2.93) & (3.32) \\

4F alpha 
& 0.0010 & 0.0022 & 0.0026*** & 0.0039*** & 0.0053*** & 0.0068*** 
& 0.0020 & 0.0010 & 0.0023*** & 0.0023*** & 0.0035** & 0.0039** \\

& (0.60) & (1.61) & (2.69) & (3.40) & (3.05) & (4.05) 
& (1.35) & (1.05) & (2.67) & (3.00) & (2.20) & (2.10) \\

\hline
\end{tabular}%
}
\end{table*}

The ensemble effectively ranks future sector performance: predicted returns, excess returns, and estimated alphas increase monotonically from the ``Bottom 5'' to ``Top 5'' portfolios. Alphas for the ``6–20'' and ``Top 5'' portfolios are positive and statistically significant, often at the 1\% level, suggesting that the strategy captures sector-level return components not explained by traditional risk factors.

Across models, estimated alphas decline as additional factors are introduced, indicating that the size, value, and momentum factors partially explain portfolio returns. However, the persistence of statistically significant alphas—particularly for the top-ranked portfolios—implies that our ensemble strategy captures incremental return predictability beyond these benchmarks.

Finally, the last column of Table~\ref{excess_returns_on_portfolios} presents results for a market-neutral strategy that goes long the ``Top 5'' and short the ``Bottom 5'' sectors. This strategy delivers highly significant alphas, comparable to the long-only ``Top 5'' strategy, further reinforcing the robustness of our ensemble-based sector ranking.

\section{Conclusion}\label{sec:conclusion}

We introduce a novel gradient-free online ensemble algorithm based on the Multiplicative Weights Update Method, which adaptively aggregates forecasts by balancing exploration and exploitation. The method achieves strong out-of-sample $R^2$ and regret bounds approaching those of an oracle with perfect hindsight.

Unlike prior gradient-based approaches \cite{Monti:18,Capo:22}, our framework is efficient, volatility-free, and suitable for real-time applications such as hyperparameter tuning and sequential forecasting.

Empirically, we demonstrate that aggregating firm-level characteristics into sector-level features enhances return predictability. The proposed ensemble consistently outperforms individual models and yields a robust and profitable sector rotation strategy, even under transaction costs and across market regimes.

Future research may incorporate macroeconomic signals to further improve forecast accuracy, building on the insights of \cite{Chris:14} in high-dimensional ML contexts.



\bibliographystyle{plain}
\bibliography{sample-base}

\end{document}